\documentclass[%
reprint,
aps,
amsmath,amssymb,
pra,
]{revtex4-2}

\usepackage{graphicx}
\usepackage{dcolumn}
\usepackage{bm}
\usepackage{siunitx}
\usepackage{hyperref}
\usepackage{todonotes}
\usepackage{soul}

\begin{document}
\newcommand{\onesnull}{$\mathrm{(5s^2)\:^1S_0}$\xspace}
\newcommand{\threesone}{$\mathrm{(5s6s)\:^3S_1}$\xspace}
\newcommand{\threepnull}{$\mathrm{(5s5p)\:^3P_0}$\xspace}
\newcommand{\threepone}{$\mathrm{(5s5p)\:^3P_1}$\xspace}
\newcommand{\threeptwo}{$\mathrm{(5s5p)\:^3P_2}$\xspace}
\newcommand{\threedthree}{$\mathrm{(5s5d)\:^3D_3}$\xspace}

\newcommand{\red}{$\mathrm{(5s^2)\:^1S_0-(5s5p)\:^3P_1}$\xspace}
\newcommand{\green}{$\mathrm{(5s5p)\:^3P_2-(5s5d)\:^3D_3}$\xspace}
\newcommand{\blue}{$\mathrm{(5s^2)\:^1S_0-(5s5p)\:^1P_1}$\xspace}
\newcommand{\ir}{$\mathrm{(5s5p)\:^3P_2-(5s4d)\:^3D_3}$\xspace}

\newcommand{\rhoq}{\hat\rho(q)}
\newcommand{\rhozq}{\hat\rho(z,q)}

\renewcommand{\figureautorefname}{Fig.}
\renewcommand{\equationautorefname}{Eq.}

\title{Sub-Doppler cooling of bosonic strontium in a two-color MOT}

\author{Milán János Negyedi}
\affiliation{Physikalisches Institut, Eberhard Karls Universität Tübingen, Auf der Morgenstelle 14, D-72076 Tübingen, Germany}
\author{Shubha Deutschle}
\affiliation{Physikalisches Institut, Eberhard Karls Universität Tübingen, Auf der Morgenstelle 14, D-72076 Tübingen, Germany}
\author{Florian Jessen}
\affiliation{Physikalisches Institut, Eberhard Karls Universität Tübingen, Auf der Morgenstelle 14, D-72076 Tübingen, Germany}
\author{József Fortágh}
\affiliation{Physikalisches Institut, Eberhard Karls Universität Tübingen, Auf der Morgenstelle 14, D-72076 Tübingen, Germany}
\author{Lőrinc Sárkány}
\affiliation{Physikalisches Institut, Eberhard Karls Universität Tübingen, Auf der Morgenstelle 14, D-72076 Tübingen, Germany}

\begin{abstract}
We report a two-color magneto-optical trap that continuously operates on the \blue and \green transitions in $\mathrm{^{88}Sr}$. Owing to the \mbox{sub-Doppler} cooling from the metastable \threeptwo state, a bimodal momentum distribution is observed, with the cold part of the ensemble reaching a temperature of \SI{105}{\micro\kelvin}. A detailed simulation is established to explain the data. The absolute frequency of the \green transition is measured to be \mbox{603 976 473 \textpm~2~MHz} using a frequency comb, improving the literature value. The accompanying magnetic trap is characterized, and is shown to contribute to the sequential production of ensembles with temperatures down to \SI{45}{\micro\kelvin}.

\end{abstract}

\maketitle

\section{Introduction}
Optical lattice clocks are a cornerstone of quantum metrology, redefining the accuracy of time and frequency standards, offering improvements in commercial applications such as GNSS systems \cite{GNSSeval, Major2014, Schuldt2021}, telecommunication \cite{intro_telecom}, chronometric levelling \cite{Grotti2018}, and earthquake forecasting \cite{2015EPJWC..9504009B}, as well as tackling fundamental problems in physics, such as testing general relativity \cite{Takamoto2020}, dark matter research \cite{Derevianko2014, PhysRevLett.114.161301}, gravitational wave detection \cite{PhysRevD.94.124043, Tino2019}, or variations in physical constants \cite{PhysRevLett.120.173001, Barontini2022, PhysRevLett.120.173001}. 

A major limitation of their performance is the Dick effect that is a direct consequence of the standard, cyclic operation principle of the clock \cite{dick, Santarelli1998}: alkaline-earth atoms typically have to be cooled on multiple transitions, subsequently, in order to reach the necessary {\micro\kelvin} temperatures, and are discarded after the interrogation. This leads to an effective downconversion of the phase noise of the clock laser into lower frequency regime, compromising the clock stability.

Various ideas have been proposed and demonstrated to mitigate this effect, such as dead-time free operation by combining two separate clock systems \cite{PhysRevLett.111.170802, Schioppo2017, Oelker2019}, non-destructive measurements \cite{Vallet2017, PhysRevA.79.061401}, or reusing the same atoms by recooling them after a spectroscopy cycle has been completed \cite{PhysRevLett.122.173201}.

Continuously operated clocks may offer a simpler alternative route to suppress the Dick effect, therefore continuous preparation of ultracold alkaline-earth atoms is of utmost importance. Recently, continuous sources of ultracold strontium atoms have been demonstrated based on the \blue and \red \cite{PhysRevLett.119.223202, Niu23}, as well the on the \ir transition \cite{Katori2023}.

In this paper we investigate the feasibility of a single-step, continuous source of ultracold strontium atoms based on a MOT operated on the \emph{green} \green transition, loaded directly from a dispenser instead of using a Zeeman-slower, reducing system size and complexity. To the best of our knowledge, we have achieved the lowest reported temperature on this transition, cold enough that the atoms can be efficiently transferred into an optical dipole trap. A detailed treatment is provided to explain the observed bimodal momentum distribution of the atoms. We have determined the absolute frequency of the \green transition using a frequency comb, improving the literature value \citep{Stellmer_reservoir}. The measured frequency is 603 976 473 \textpm~2~MHz.

Furthermore, the linear magnetic trap created by the MOT quadrupole coils is investigated, and is shown to aid in the sequential generation of atomic ensembles down to \SI{45}{\micro\kelvin}, with populations of more than 20 thousand atoms.

\section{Two-color MOT}
Using the \emph{green} \green strontium transition for cooling has been considered several times in the literature \citep{Stellmer_reservoir, Krutzik_laser}. The idea behind it is to benefit from electrons being shelved into the metastable \threeptwo state while the \emph{blue} MOT on the \blue transition is being operated, allowing concurrent operation of the blue and green cooling transitions, lowering the temperature of the trapped atoms. Additionally, since the \threeptwo state, in contrast to the \onesnull ground state, does have quasi-degenerate substates, sub-Doppler cooling \citep{Cohen-Tannoudji} is possible to occur even for the bosonic Sr isotopes, enhancing cooling even further.
\begin{figure}[!h]
    \centering
    \includegraphics[scale=0.95]{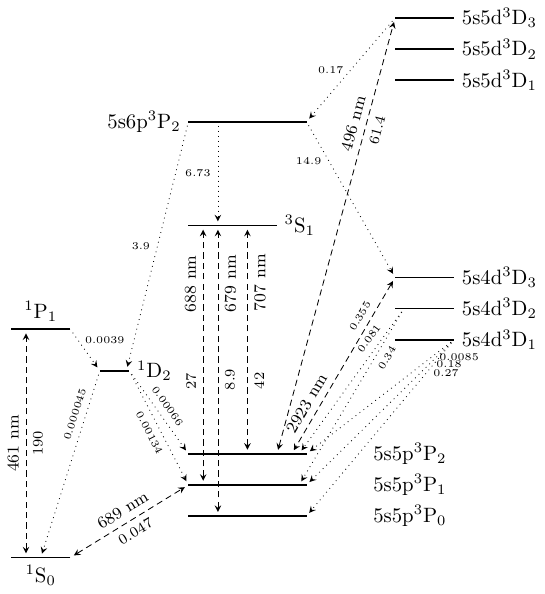}
    \caption{The term scheme for Sr. Transition and decay rates are given in units of $\mathrm{10^6/s}$. \citep{Katori_greenmot}}
    \label{fig:level}
\end{figure}

The green transition is not completely cycling, atoms are lost to the \threepone state through various channels, as seen in Fig.~\ref{fig:level}, therefore it is unfavorable for laser cooling \citep{Stellmer_reservoir} in itself. The \emph{mid-infrared} or \emph{MIR} \ir transition does not suffer from the same problem (it is completely closed), therefore it was considered the preferable alternative, especially given the much lower Doppler ($T_D$) and recoil ($T_r$) temperature limits: $\mathrm{T_{D,g}=\SI{230}{\micro\kelvin},T_{r,g}=\SI{0.23}{\micro\kelvin}}$, in contrast to $\mathrm{T_{D,MIR}=\SI{1.4}{\micro\kelvin}, T_{r,MIR}=\SI{0.013}{\micro\kelvin}}$ \citep{Katori_greenmot}.
The \SI{2.9}\micro\meter~wavelength of the mid-infrared transition presents its own unique challenges however, and so far there are only a few reports on the realization of such a MOT \cite{Hobson, Katori_greenmot}.

Recently, it has been demonstrated that with some modification of the cooling scheme, the green transition is indeed capable of creating an ultracold sample of strontium atoms \citep{Katori_greenmot}.

Using the transition to the $5s5d$ triplet in comparison to $5s4d$ comes with several advantages. First, the wavelength (\SI{496.3}{\nano\meter}) is in the visible range, and while it is currently somewhat difficult to come by, direct laser diodes are beginning to appear on the market. The general availability of optical components is better compared to the mid-infrared wavelength, and it also makes the adjustment of optical paths significantly easier. Second, the green transition is about 2.5 orders of magnitude broader than the MIR one, drastically reducing the complexity of the necessary laser frequency stabilization techniques.

\subsection{Experimental setup}
\label{sec:ExpSetup}
\begin{figure}[!h]
    \centering
    \includegraphics[scale=0.35]{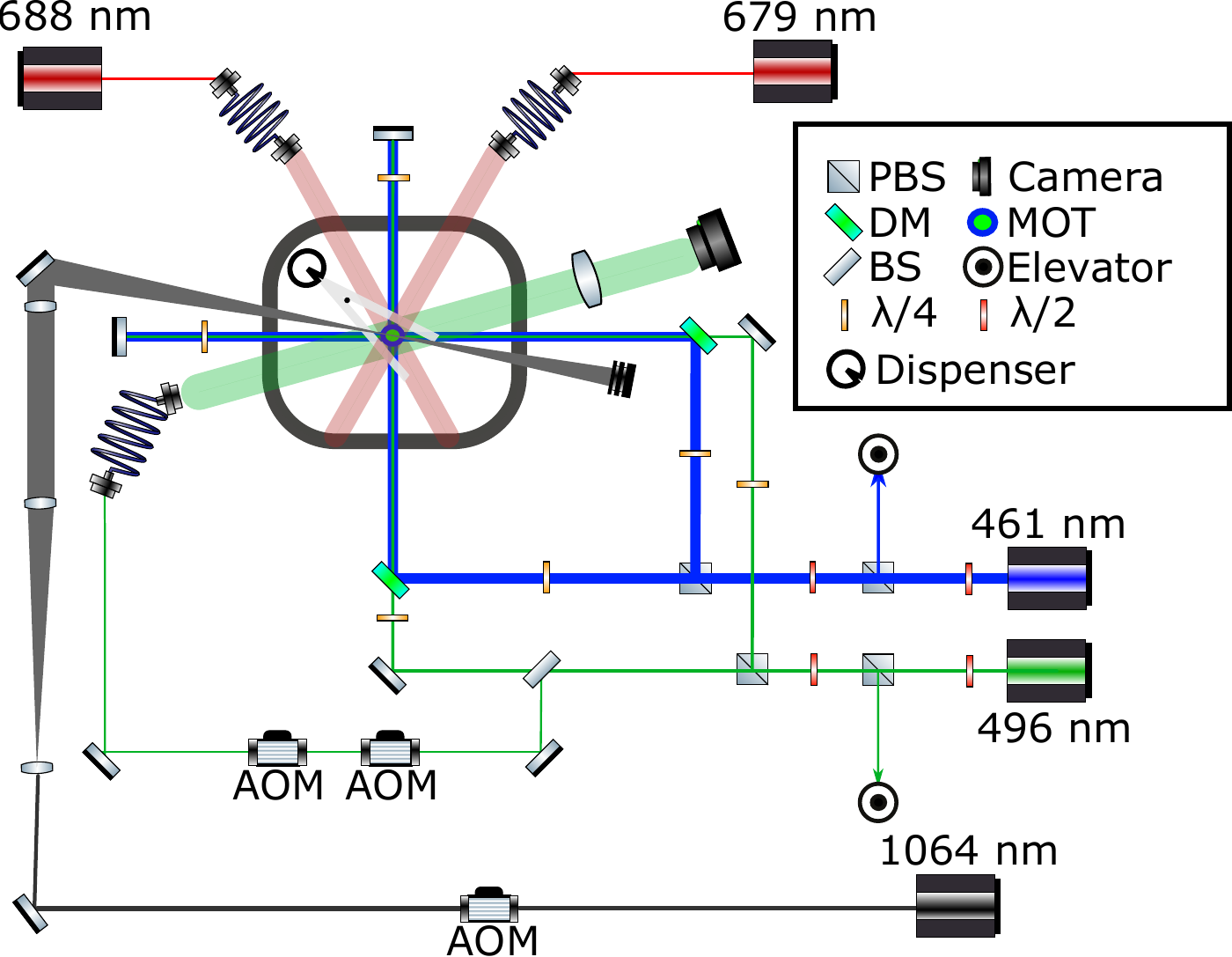}
    \caption{A diagram of the general optical arrangement of the experiment. Some beam guiding mirrors were omitted for the sake of clarity. PBS: polarizing beamsplitter, DM: dichroic mirror, BS: beam sampler. The elevators create the vertical cooling beams.}
    \label{fig:opt_setup}
\end{figure}

The optical setup is shown in Fig.~\ref{fig:opt_setup}. The two-color MOT is operated in an overlapped, retroreflected configuration. The cooling beams are first split using polarizing beam splitter cubes, then the polarization of each beam is controlled with achromatic quarter wave plates. The respective blue and green beams are then overlapped concentrically on dichroic mirrors and the three beams are sent into the chamber in a pairwise orthogonal arrangement. On the other side of the chamber they pass through an achromatic quarter wave plate and then are reflected back by a mirror. The $\mathrm{1/e^2}$ beam diameters are 12 and 4.5 mm for the blue and green beams, respectively.

The atomic cloud is detected by absorption imaging using an achromatic doublet in a 2f-2f configuration. The imaging light was derived from the green cooling beam using a beam sampler, and was shifted back 19~MHz to resonance using two AOM-s before entering the chamber.

The blue (\SI{461}{\nano\meter}) cooling laser is frequency doubled from a \SI{922}{\nano\meter} tapered amplifier (TA), using a PPLN crystal (NTT Electronics), capable of delivering a maximum cooling power of 40 mW.
The TA is seeded by a grating-stabilized ECDL. The green (\SI{496}{\nano\meter}) setup is analogous, with a maximum cooling power of 16 mW.
The repumping lasers (679 and \SI{688}{\nano\meter}) are grating-stabilized ECDLs, with maximum powers of 10 and 3 mW incident on the vacuum chamber windows, respectively.
The cooling lasers are delivered free-beam to the optical setup, whereas the repumpers are transmitted through singlemode optical fibers \citep{Lorinc_phd}. All lasers are frequency stabilized using a wavemeter (HighFinesse WS8-2).

The experiment is operated in a small spherical vacuum chamber \citep{Lorinc_phd} at a base pressure of $\mathrm{5\cdot10^{-11}}$ mbar. A pair of in-vacuum coils oriented along the horizontal axis running opposite currents generates a magnetic quadrupole field with a gradient of 57 G/cm along the symmetry axis of the coils, which we denote as the \emph{z-axis} in this work. 
The chamber facilitates six viewports for the cooling beams, while the repumpers and imaging beams are placed at arbitrary angles in the horizontal plane. The top viewport has a larger diameter allowing observation of fluorescence. All instrumentation (ion getter pump, electrical feedthroughs, pressure gauge) is placed below the science chamber to minimize conflicts with the optics and to simplify installation. The atom source (dispenser) is attached to the outer circumference of the coils emitting strontium towards the central area of the quadrupole. The dispensers are operated continuously, increasing the pressure to $\mathrm{4\cdot10^{-10}}$ mbar. The hot atoms are prevented from directly hitting atoms in the MOT volume by a shielding wire.

According to the literature, this setup is the first one where direct loading of a 3D-MOT from a dispenser without any kind of precooling was successfully implemented for strontium. This significantly reduces experimental complexity, allowing for a compact design, and paves the way for miniaturized, transportable systems. When using the green transition as a repumper, the resulting \emph{blue} MOT can be routinely loaded with 1.5 million atoms, we measure an atomic flux of $\Phi_{disp} = 3.0\cdot10^6 \pm 1.5 \cdot 10^5~1/\mathrm{s}$ with a two-body loss rate of $\beta_b=1.65\cdot 10^{-6} \pm 3 \cdot 10^{-8}~1/\mathrm{s}$.

\subsection{Characterization of the green MOT}
Owing to the degeneracy of the \threeptwo state, \textit{$\mathrm{\sigma^{+} - \sigma^{-}}$} sub-Doppler cooling is taking place concurrently upon operating the green MOT, as confirmed by a recent experiment \cite{Katori_greenmot}. Numerical simulations have shown that for an arbitrary set of parameters typically neither the density, nor the momentum of the atoms will follow a Gaussian  distribution \citep{dual_dist1, dual_dist2, dual_dist3, dual_dist4}. During the measurements presented below, the \emph{blue} cooling beam intensity was kept at 9 $\mathrm{mW/cm^2}$, and its detuning at 51 MHz. The repumper beams were kept at maximum intensity and on resonance. All intensities given are single-beam intensities. All presented data refers to the green MOT and the atomic ensemble it generates, unless specifically noted otherwise.

\begin{figure}[!h]
%    \centering
    \includegraphics[scale=1]{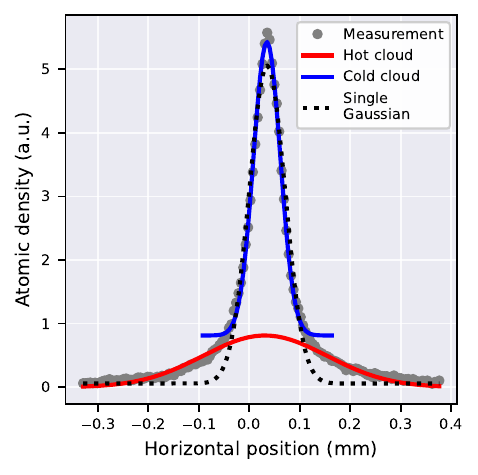}
    \caption{Measured horizontal density distribution in the green MOT, and a comparison between fitting with one or two Gaussians. $\mathrm{I/I_{sat}= 0.4,\:\Delta = 1.9\,\Gamma}$, G=57~G/cm}
    \label{fig:green_dist}
\end{figure}
The density distribution of the cloud, integrated over the vertical axis, is plotted in Fig.~\ref{fig:green_dist}\, for detuning $\Delta=1.9\,\Gamma$ and a $G=57$~G/cm magnetic gradient along the \mbox{\emph{z-axis}}. It does not follow a simple Gaussian distribution: the atoms participating in the green cooling cycle separate into two distinct, overlapping clouds, suggesting two distinct components of the velocity distribution that we will label as the \emph{cold} and the \emph{hot} cloud. Therefore, we fit the measured atomic density profile by the sum of two Gaussians, and from this we determine the corresponding velocity distributions $F_c(v)$ and $F_h(v)$ by the usual ballistic expansion method. The velocity distribution of the entire cloud $F(v)$ is calculated as the weighted average of $F_c(v)$ and $F_h(v)$, using the respective fitted atomic populations as weights. The effective temperature of the entire cloud is defined through the expectation value of the squared velocity under $F(v)$:

\begin{equation}
T_\text{eff} = \frac{m_{Sr} \left\langle v^2 \right\rangle_{F}}{k_B}.
\end{equation}

\begin{figure}[!h]
    \centering
    \includegraphics[scale=1]{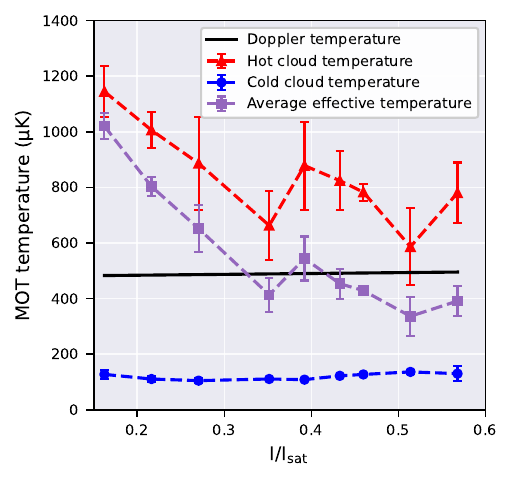}
    \caption{Measured temperatures of the \emph{hot} (red) and \emph{cold} (blue) clouds, and the calculated effective temperature of total green MOT population as a function of green cooling laser intensity. $\mathrm{\Delta = 1.9\,\Gamma, G=57~G/cm}$}
    \label{fig:green_I}
\end{figure}

\begin{figure}[!h]
\centering
  \includegraphics[scale = 1]{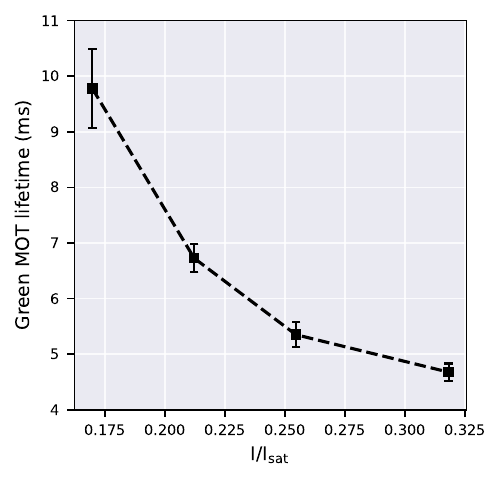}
  \caption{The lifetime of atoms in the green cooling cycle as a function of green cooling laser intensity. $\mathrm{\Delta = 1.9\,\Gamma, G=57~G/cm}$}
  \label{fig:life}
\end{figure}

\begin{figure}[!h]
  \centering
  \includegraphics[scale = 1]{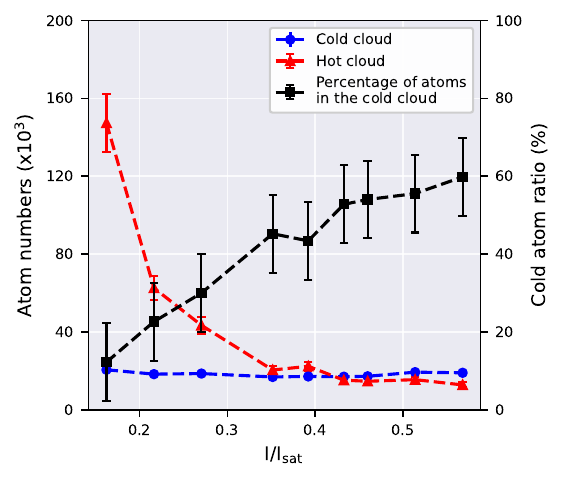}
  \caption{Atom numbers in the \emph{hot} (red) and \emph{cold} (blue) cloud, and fraction of the cold atoms compared to the total atom number in the green MOT (black) as a function of green cooling laser intensity. $\mathrm{\Delta = 1.9\,\Gamma, G=57~G/cm}$}
  \label{fig:green_num}
\end{figure}

The effective temperature of the MOT is plotted as a function of the cooling intensity in Fig.~\ref{fig:green_I}, along with the temperature of the cold and hot fractions of the cloud, and the Doppler limit of the green transition. The lifetime of atoms in the green cooling cycle of the MOT was measured by turning off the blue cooling beams, thereby preventing it from loading. Although direct capture of a very small quantity of atoms into the green cooling cycle is theoretically possible, it is below the detection limit of our system, and was therefore neglected. The data is shown in \autoref{fig:life}, while the populations of the two clouds and the ratio of the sub-Doppler cooled atoms are shown in \autoref{fig:green_num}.

The MOT consistently produces a cold population with a temperature below \SI{140}{\micro\kelvin}, well below the Doppler limit of the green transition for a broad range of parameters. The coldest temperature reached is 105 $\mathrm{\pm}$ 9 \SI{}{\micro\kelvin}. This temperature is comparable to \cite{Katori2023}, despite the fact that a different transition is used, and is sufficiently low so that atoms can be efficiently loaded into an optical dipole trap or to a moving optical lattice. We have in fact loaded up to $9\times10^3$ atoms into a dipole trap operated at \SI{1064}{\nano\meter} with \SI{30}{\micro\meter} beam waist and \SI{5}{\watt} of power, and measured a lifetime of \SI{600}{ms}. This is in good agreement with the lifetime measured in the magnetic trap (see Sec.~\ref{sec:MagneticTrap}), confirming that it is limited by background gas and Sr collisions.

\begin{figure}[!h]
    \centering
    \includegraphics[scale=1]{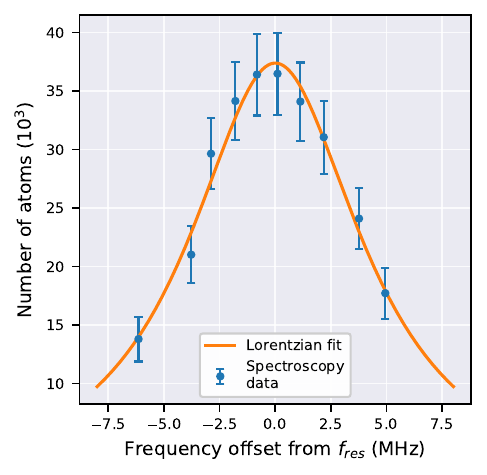}
    \caption{Absorption spectroscopy of the green transition. $\mathrm{f_{res} = 603\,976\,473 \pm 2\SI{}{\mega\hertz}}$}
    \label{fig:dip_spec}
\end{figure}

We have determined the absolute frequency of the \green transition by performing absorption spectroscopy on the cold atoms, as shown in Fig.~\ref{fig:dip_spec}. The cooling lasers and the magnetic coils were turned off \SI{1}{ms} before imaging to eliminate Zeeman and light shifts. There are no eddy currents \SI{500}{\micro s} after the trigger signal for turning off the coils, as confirmed by a current clamp.

The frequency of the fundamental mode of the cooling laser has been measured by beating it with a frequency comb (Toptica DFC CORE+ with DFC SCIR extension module, \SI{80}{\mega\hertz} repetition rate) that is referenced to a GPS-disciplined rubidium frequency standard (Precision Test Systems GPS10RBN).

The comb has an absolute frequency error of $\mathrm{\sim}$ \SI{3}{\kilo\hertz}, limited by the instability of the RF reference. The AOM-s used to shift the green frequency back towards resonance are controlled by a waveform generator (Rigol DG4102), while the beat frequency was measurement by a spectrum analyzer (Signal Hound BB60C). Considering the error of the Lorentzian fit, the various instrument errors and the Zeeman shift resulting from the Earth's magnetic field, we estimate the total absolute error to be less than 2 MHz. As such, we determine the transition frequency to be \mbox{$\mathrm{f_{res} =603\,976\,473 \pm 2}$ \SI{}{\mega\hertz}}, in slight disagreement with \citep{Stellmer_reservoir}.	

\subsection{Quantum treatment of sub-Doppler cooling}
Following \cite{prudnikovQSubDop}, we have performed detailed numerical simulations to provide a theoretical explanation of the observed momentum distribution and effective temperature. We consider a one-dimensional problem along the $z$ axis with two counterpropagating laserbeams with $\sigma^+/\sigma^-$ polarizations, respectively. 

\begin{figure}[!ht]
    \centering
    \includegraphics[scale=1]{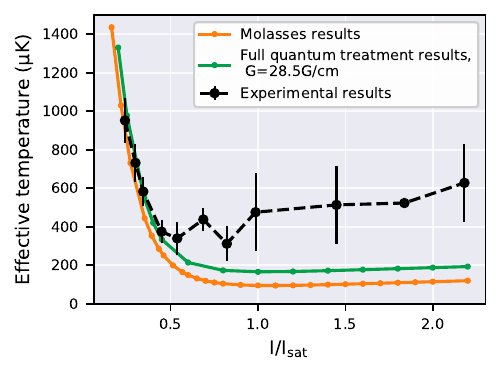}
    \caption{The results of the full quantum treatment for a molasses and for a MOT compared with the experimental effective temperature results, as a function of green cooling laser intensity}
    \label{fig:subdop_comp}
\end{figure}

As a first approximation, we neglect the effects of the magnetic fields in the MOT, considering only optical molasses. Working in two-point coordinate representation $\hat\rho(z_1,z_2)=\langle z_1|\hat\rho|z_2\rangle$, and changing into the coordinate system rotating along with the polarization of the total light field, the time evolution of the density matrix is described by the following equation:

\begin{equation}
\begin{split}
	2\omega_\text r\,\partial_q [\hat J_\text z, \rhoq] = &\hat\Gamma\left\{\rhoq\right\} + i\Delta\left[\hat P^e, \rhoq\right]+\\
	&i\Omega\left(\hat V_1(q)\rhoq - \rhoq \hat V_2(q)\right),
\end{split}
\end{equation}
subjected to normalization $\text{Tr } \hat\rho(q=0)=1$, where \mbox{$q= k(z_1-z_2)$} is the dimensionless relative coordinate, $k=2\pi/\lambda$, $\omega_r = \hbar k^2 / 2M$ with $M$ the atomic mass, $\hat\Gamma$ is the spontaneous atomic relaxation operator, as defined in Eq.~(16) and Eq.~(17) in \cite{prudnikovQSubDop}, and $\Omega\hat V_1(q)=\Omega\hat V_2^\dagger(q)$ is the atom-light interaction with Rabi frequency $\Omega$, $\Delta$ is the detuning, and

\begin{equation}
V_1(q)=\hat T_{-1}\,\text e^{-iq/2} - \hat T_{+1}\,\text e^{iq/2} + \text{h.c.},
\end{equation}
with
\begin{equation}
\hat T_\sigma=\sum_{m_e, m_g} C^{J_e, m_e}_{J_g, m_g; 1\sigma}\,|J_e,m_e\rangle\langle J_g,m_g|.
\end{equation}
The results are depicted in \autoref{fig:subdop_comp}. 

Although this simple optical molasses calculation does already explain the non-Gaussian distribution of momentum, it fails to exactly reproduce the measured dependence on cooling power. Similarly to \cite{dual_dist1}, this can be attributed to the fact that the effect of the magnetic field in the MOT can only be neglected if 

\begin{equation}
\frac{1}{\hbar}\,\mu_B g_J B < \frac{\Gamma s}{1+s + \left(\frac{2\Delta}{\Gamma}\right)^2}\,,
\end{equation}
a condition which is violated in our case.

To estimate the effect of the magnetic field, we have followed the method outlined in \cite{dual_dist1}. We note that for the parameter range we have covered, the movement of the atoms in the MOT can be considered adiabatic, i.e. the trap frequency is much smaller than the cooling rate, therefore at a given position $z$ the equilibrium momentum distribution is determined by the local magnetic field. This allows us to approximate the total momentum distribution as a weighted average of the partial contributions originating from the subensembles of the cloud distributed along the $z$ axis, weighted by the experimentally determined density distribution. The results of this calculation are shown in \autoref{fig:subdop_comp} under "Full quantum treatment".

The approximative calculation correctly predicts that the plateau of the effective temperature occurs at a higher temperature than for the molasses (\autoref{fig:subdop_comp}). The optimal density and temperature of the cold cloud is reached for a saturation parameter between 0.25-0.5 (see \autoref{fig:green_I}) where this model already shows reasonable agreement with the measurements. For higher intensities the discrepancy between the measured and the predicted temperatures is more pronounced, suggesting that a self-consistent quantum treatment is necessary to explain the observed data. This will be the subject of a future work.

\section{Magnetic trap}
\label{sec:MagneticTrap}
In the absensce of cooling light, the quadrupole field of the MOT coils forms a linear magnetic trap (MT) for the low-field seeking Zeeman substates of the \threeptwo state. Typical gradients used for the MOT ($45-60$~G/cm) are sufficient to hold the atoms against gravity \citep{magtrap}. In our experiment, we have investigated the loading of the sub-Doppler cooled ensemble into the magnetic trap by turning off the cooling beams and observed a reduction of the temperature after the atoms have been transferred to the MT. Transfer efficiency of the cold cloud atoms to the MT was 60 percent. We have measured a lifetime of $\tau\approx0.52$~s, limited mainly by collisions with background gas and hot Sr atoms.
The cooling effect upon transferring to the magnetic trap could open up additional possibilities for continuous preparation of the atoms based on the principles outlined in \citep{mag1, mag2, mag3}, using the magnetic field to transport and buffer the sequentially produced cold atoms for continuous loading into an optical lattice.

Due to the low atom number and shallow magnetic potential, at flight times higher than 1 ms the signal-to-noise ratio is prohibitively low for using the standard ballistic expansion method. This is compounded by the fact that the starting distribution of the MT is not Gaussian. Therefore we have used the technique described in \citep{magtrap} to evaluate the temperature of the cloud instead. Having calibrated the magnetic field of the coils allows us to calculate the density distribution in the trap potential, with only the temperature and the atom number as fitting parameters. Integrating the distribution along the axis of the imaging beam ($y$), we obtain the fitting function:
\begin{equation}
F(x,z,T,n_0) = n_0 \cdot \text{exp}\left(-\frac{m_J \mu_B g_J B(x,z) - m g z}{k_B T}\right),
\end{equation}
where $g_J=3/2$ for the \threeptwo state, and $n_0$ is the 2D optical density amplitude.

\begin{figure}[!h]
    \centering
    \includegraphics[scale=1]{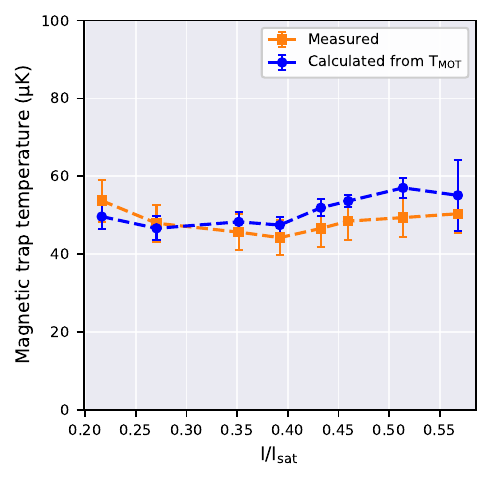}
    \caption{$T_\text{MT}$ as a function of green cooling beam intensity, as determined from absorption imaging (squares) and from $T_\text{MOT}$, using Eq. \ref{eq:mag} (circles)}
    \label{fig:mag_comp}
\end{figure}

Since the magnetic trap is populated mainly from the cold cloud, overwhelming majority of the atoms will be in the $m_J = +2$ state, therefore we have assumed $m_J = +2$ across the whole sample. Using this method, we consistently measure temperatures $T_{MT}$ between 45 and \SI{60}{\micro\kelvin} in the magnetic trap, largely independent of the MOT beam intensity, significantly colder than the cold cloud of the MOT, see Fig.~\ref{fig:mag_comp}. The lowest temperature in the MT we have achieved is \SI{44(5)}{\micro\kelvin}. According to the available literature, this is the lowest value reported for strontium in a magnetic trap loaded from a MOT.

The temperature change when transferring atoms to the MT from the MOT can be estimated in an independent way using the virial theorem \citep{stuhler_magtrap}: since the MT potential is linear, the virial theorem ensures that $V_{MT}=2E_{MT}$, where $E_{MT}$ and $V_{MT}$ is the average kinetic and potential energy of the atoms in the MT, respectively. Assuming a negligible-sized MOT would lead to $E_{MOT}=E_{MT}+V_{MT}=3E_{MT}$, i.e. $T_{MT} = T_{MOT}/3$. The finite size of the MOT slightly modifies this result to

\begin{equation}
\label{eq:mag}
T_{MT} = \frac{T_{MOT}}{3} + \frac{8}{9\sqrt{2 \pi}} \frac{\mu_b g_J m_J}{k_B}\,G \sigma,
\end{equation}
where $\sigma$ is the radius of the MOT. The results from the two methods are in agreement with each other within measurement error, as shown in Fig.~\ref{fig:mag_comp}

\section{Conclusions and outlook}

In this paper we have demonstrated the creation of a green strontium MOT operated on the \blue and \green transitions, and provided a detailed theoretical and experimental analysis of the emerging momentum distribution caused by sub-Doppler cooling processes. We characterized the accompanying magnetic trap, and investigated its additional cooling effects. Our system promises to be an excellent, easy-to-realize platform to continuously source ultracold atoms with suffient flux and temperature to be loaded into a moving optical lattice, paving the way towards continuously operated clocks and superradiant lasers.

\begin{acknowledgments}
We thank H. Katori for for the joint initiation of the Strontium project, and N. Ohmae for technical consultation. The authors gratefully acknowledge financial support from the Baden-Württemberg Stiftung under project number BWST ISFIII 15, and valuable discussion with O. Prudnikov on the theoretical framework of sub-Doppler cooling. We thank C. Groß for the loan of the dipole trap laser. S Deutschle and M. Negyedi thank A. Günther for helpful discussion and advice. S. Deutschle gratefully acknowledges funding from the Vector Stiftung via project number P2023-0105.
\end{acknowledgments}
\newpage
\bibliography{refs}

\end{document}